\documentclass[11pt,a4paper]{article} 
\usepackage{jcappub} 
\usepackage{mathrsfs}
\usepackage{bbm}
\usepackage{amsfonts}
\usepackage{graphicx}
\usepackage{natbib}

\newcommand{\refeq}[1]{(Eq.\,\ref{eq-#1})}
\newcommand{\fig}[4]{\begin{figure}[htbp]\centering\includegraphics[width=#3\textwidth]{#1}\caption{#2}\label{fig-#1}\end{figure}}
\newcommand{\refig}[1]{Fig.\ \ref{fig-#1}} 
\usepackage{hyperref}
\usepackage{color} 

\def\lsim{\, \rlap{$<$}{\lower 1.1ex\hbox{$\sim$}}\,}

\newcommand{\void}[1]{}

\title{The Scale-invariant Power Spectrum of Primordial  Curvature Perturbation in CSTB Cosmos} 
\author[]{Changhong~Li ,} 
\author[]{Yeuk-Kwan~E.~Cheung
} 
\affiliation[]{Department of Physics, Nanjing University,\\ 
22 Hankou Road, Nanjing, China 210093} 
\emailAdd{chellifegood@gmail.com} 
\emailAdd{cheung@nju.edu.cn} 

\abstract{%
We investigate the  spectrum of  cosmological perturbations in a bounce cosmos modeled by a scalar field coupled to the string tachyon field (CSTB cosmos). 
By explicit computation of  its primordial spectral index   
we show  the power spectrum of curvature perturbations, generated during the tachyon matter dominated contraction phase, to  be  
nearly scale invariant. 
We propose a unified space of parameters for a systematic study of inflationary/bouncing cosmologies. 
We find that CSTB cosmos is dual--in Wands's sense--to the slow-roll inflation model as can be easily seen from this unified parameter space. 
Guaranteed by the dynamical attractor behavior of CSTB Cosmos, this scale invariance is free of the fine-tuning problem, in contrast to the slow-roll inflation model. 
} 
\arxivnumber{1234.5678} 
\begin{document} 
\maketitle

\section{Introduction}
\label{sec:intro}

In accordance with observations of Cosmic Microwave Background 
(CMB) anisotropies~\cite{Komatsu:2010fb,Planck:2013kta} the scale-invariance of power spectrum of the primordial curvature perturbations serves as a crucial criterion for testing the validity of early-universe models. A well-known example is the exponential inflation driven by a slowly-rolling scalar field, the spectrum of curvature perturbations of which could be nearly scale-invariant by fine-tuning the flatness of the inflaton potential. However--besides this fine-tuning problem of the flatness of potential\footnote{With a slight change of the potential of the scalar field, the background is no longer an exponential expansion. And the spectrum of curvature perturbation becomes time-dependent, which in turn renders it  scale-variant implicitly~\cite{Liddle:2000cg, Dodelson:2003ft}.}--slow-roll inflation suffers another severe problem: the singularity of its initial conditions, {\it i.e.} the  Big Bang Singularity~\cite{Borde:1993xh}.

Although proven challenging  many  bounce/cyclic universe models%
%%%%%%%%%%%%%
\footnote{The idea of a collapsing phase preceding a phase of expansion could be traced back to three giants Tolman, Einstein and Lemaitre who, independently, proposed 
the idea in the early 1930s, for example see~\cite{Tolman:1931fi}.} 
%%%%%%%%%%%%%
have been proposed  in an effort to address  the  problems of the Big Bang Singularity~\cite{Novello:2008ra,Brandenberger:2012zb,  Bassett:2005xm, Lehners:2011kr, Khoury:2001wf, Khoury:2001bz, Cai:2007qw, Cai:2009zp, Gasperini:2002bn, Creminelli:2006xe, Creminelli:2007aq, Abramo:2007mp}. Moreover, to get a scale-invariant curvature spectrum in bounce universe scenario, Wands introduced a remarkable mechanism in~\cite{Wands:1998yp} (see also~\cite{Finelli:2001sr, Durrer:1995mz}): the scale invariant spectrum of a single scalar field perturbation can--seemingly--be generated during a matter-dominated contraction phase. 
The idea of Wands has since been warmly embraced in many variations on the theme of the matter-bounce universe models~\cite{Brandenberger:2009yt, Peter:2008qz, Lidsey:1999mc, Cai:2009in, Allen:2004vz, Wang:2009rw, Cai:2008qw, Cai:2011zx, Lin:2010pf, Piao:2009ku, Cai:2012va, WilsonEwing:2012pu}. 
However, as it was first pointed out by~\cite{Gratton:2003pe}, the perturbation modes grew out of the horizon indicating an unstable cosmological background.

In order to achieve a physically scale-invariant perturbation spectrum which does not renders its own cosmological background unstable, in either inflationary or cyclic/bounce cosmos, 
we extended--and analyzed in detail--the parameter space governing  the  equations of motion of the cosmological perturbations~\cite{Li:2012vi}. 
With the standard assumption that Equation of State(EoS) of the cosmological background being constant in the period of  perturbation generation\footnote{Relaxing the constant EoS assumption, however, the scale-invariant power spectrum can  also be generated in a slowly contracting Ekpyrotic background~\cite{Khoury:2009my, Khoury:2010gw, Joyce:2011ta}. See~\cite{Linde:2009mc} for critiques of this category of models.},  
the scale factor of the cosmological background is a power law in conformal time, $a\propto \eta^\nu$. 
On other hand, relaxing the conventional assumption that the universe must be driven by one single canonical scalar field\footnote{There are many extensions for non-canonical and/or multi-fields cosmological models, for example, see~\cite{Garriga:1999vw, Senatore:2010wk, Langlois:2008qf, Shiu:2011qw}.},
%%%%%%%%%%
the Hubble friction term (or the red/blue-shift term as it is sometimes called) in the equation of the perturbation mode becomes $mH\dot{\chi}_k$ and $m$ becomes a free parameter\footnote{For the single canonical scalar field models we always have $m=3$, where $3$ comes from the spatial dimensions of our presently observable universe. 
However, in the non-canonical single/multi-field cases,  
$mH=3f(\phi_i,\dot{\phi}_i)H$ generically with $f$ being a function determined by the underlying models. 
For instance, in the tachyon field model, one gets $mH=3\sqrt{1-\dot{T}^2}H$. 
And in the  tachyon matter condensation phase, 
$\dot{T}\rightarrow 1$, {\it i.e.} 
$f(T,\dot{T})=\sqrt{1-\dot{T}^2}\rightarrow 0$, 
so that we have $m$ approach $0$ rather than $3$ in this case.  Without loss of generality we can for the time being take $m$ to be constant for any given model.}. 
In this framework  different values of $\nu$ and $m$ indicate, respectively,  different cosmological background and different Hubble friction terms in the equations of perturbations. 
Therefore, for one given model, it can be characterized by a point, $(\nu, m)$ , in the $\nu$--$m$ parameter space. 
In particular,  $(\nu, m)=(-1,3)$  for a slow-roll model, 
$(\nu, m)=(2,3)$ for Wands's matter-dominated contraction~\cite{Wands:1998yp, Finelli:2001sr, Durrer:1995mz} , 
and $(\nu, m)=(2,0)$ for the Coupled-Scalar-Tachyon Bounce 
model~\cite{Li:2011nj}. 
%%%%%%%%%%%

In general there are two groups  of scale-invariant solutions generated in an expanding or a contracting background, as tabulated in Table~\ref{table1}. 
Outside of the horizon,  solutions belong to Group I and Group II are 
stable while those in Group III and  Group IV grow and render the  background unstable~\cite{Li:2012vi}. 
It is easy to check that  the slow-roll inflation, $(\nu, m)=(-1,3)$, belongs to Group I--scale-invariant and stable solutions  in an expanding background.  
And Wands's model of matter dominated contraction, $(\nu, m)=(2,3)$, belongs to Group III which is scale-invariant but not stable in  a contracting background. 
It is worth emphasize that a physically acceptable bounce universe model with its spectrum of density perturbation being generated in the  pre-bounce contraction  should be scale-invariant and stable and {\it i.e.} satisfying the conditions $m = -\frac{2}{\nu}+1$, $\nu>0$ of Group II.

\begin{table}[htdp]
\caption{Four groups of scale invariant solutions in the $(\nu, m)$ parameter space.}
\label{table1}
\begin{center}
\begin{tabular}{|c|c|c|}
\hline
& Stable & Unstable \\

\hline
Expanding Phase &  I: $m = -\frac{2}{\nu}+1$, $\nu<0$  &  
IV: $m = \frac{4}{\nu}+1$, $\nu<0$ \\
&  Example:  & Example: \\
& Slow-roll inflation $(\nu,m)=(-1,3)$   & Unknown yet \\
\hline
Contracting Phase &  II: $m = -\frac{2}{\nu}+1$, $\nu>0$ & III: $m = \frac{4}{\nu}+1$, $\nu>0$\\
 &  Example: CSTB $(\nu,m)=(2,0)$ & Example: Wands $(\nu,m)=(2,3)$ \\
\hline

\end{tabular}
\end{center}
\label{default}
\end{table}%

In this paper  we undertake a thorough study of the cosmological perturbations  generated in the recently proposed  Coupled-Scalar-Tachyon Bounce (CSTB) model~\cite{Li:2011nj}. 
As a bounce universe model  
the spectrum of density  perturbations in CSTB cosmos  is generated during its contracting phase before the bounce point; 
the spectrum of density perturbations is assumed to be unperturbed throughout the non-singular bounce as well as  its re-entry in recent expansion phase. 

In the pre-bounce contraction, CSTB cosmos enjoys the following two properties
\begin{itemize}
\item $\nu=2$: The pre-bounce contraction is dominated by the tachyon matter which behaves like cold dust. Thus we have $\omega=0$ and $\nu=2$ in  this  phase of tachyon matter dominated contraction;
\item $m= 0$: The Hubble friction term in the equation of tachyon  perturbations is that $mH=3\sqrt{1-\dot{T}^2}H$. During the tachyon matter dominated contraction the tachyon field has already condensed  and $\dot{T}^2\rightarrow 1$, therefore, we have $m=3\sqrt{1-\dot{T}^2}\rightarrow 0$~.
\end{itemize}
In other words, the tachyon field perturbations in the CSTB cosmos correspond to a point, $(\nu,m)=(2, 0)$, in the $(\nu, m)$ parameter space. This solution clearly belongs to Group II. 
It  indicates that the power spectrum of density perturbations due to the tachyon field is scale-invariant and stable. 
Furthermore during the pre-bounce contracting phase--in  which the perturbations are outside of the horizon--the spectrum of curvature perturbation, in the long wavelength limit, is related to the spectrum of tachyon perturbation by a factor $(H/\dot{T}_c)^2$, with $T_c$ being the vacuum expectation  value of the tachyon field 
in this phase.  Because of the  following characteristic
\begin{itemize}
\item $\dot{T}_c\propto H$: During the contracting phase with  tachyon matter domination  the vacuum expectation  value of  tachyon field is proportional to the  number of e-foldings of the background, $T_c\propto N\equiv \int H dt$. 
It follows that  the factor relating the spectrum of curvature perturbations and that  of the tachyon  perturbations  is a constant, $(H/\dot{T}_c)^2 = \kappa^2$,
\end{itemize}
we conclude that the curvature spectrum of CSTB cosmos is also 
scale-invariant and stable.

This paper is organized as follows: in Section 2 we review the cosmology of CSTB model; in Section 3 we calculate the primordial spectral index of curvature perturbation of CSTB cosmos, and show that its power spectrum of curvature perturbation is nearly scale-invariant and stable. 
We discuss the ``dualities''\footnote{Following the same abuse of language in the bounce literatures  by  which it merely indicates the possible  existence of a scale invariant spectrum obtained from models other than slow roll inflation.}  between slow-roll inflation model, Wands's  matter-dominated contraction and our CSTB cosmos. 
We compare their stability properties  in Section 4, and close with a conclusion and prospects in  Section 5.

%%%%%%%%%%%%%%%%%%%%
%%%%%%%%%%%%%%%%%%%
\section{Cosmological Background in CSTB Cosmos}

In this section, we review the cosmology of a string-inspired bounce universe model  driven by a canonical scalar field coupled with a tachyon field, for short the CSTB cosmos~\cite{Li:2011nj}.  
The  Lagrangian density is comprised of  three parts,
\begin{equation} \label{eq-totl}
\mathcal{L}(\phi, T)=\mathcal{L}(T)+\mathcal{L}(\phi) -\, \lambda\phi^{2} T^{2}~,  
\end{equation}
where 
$\mathcal{L}(\phi)$ is the  Lagrangian for a massive 
(no further assumption on the value of the mass is made) 
canonical  scalar field,
\begin{equation} \label{eq-sl}
\mathcal{L}(\phi)=-\frac{1}{2}\partial_\mu\phi\partial^\mu\phi-\frac{1}{2}m_\phi^2\phi^2~.
\end{equation}
The dynamics of the tachyon field is governed by
\begin{equation}
\mathcal{L}(T)=-V(T)\sqrt{1+\partial_\mu T\partial^\mu T}~,\quad V(T)=V_0\left[\cosh\left(\frac{T}{\sqrt{2}}\right)\right]^{-1}~, \label{eq-tl}
\end{equation}
describing  the annihilation process of a pair of D3--anti-D3 branes~\cite{Sen:2002in,Sen:2002nu,Gibbons:2002md}. In effective string theory, $\phi$ can simply  be viewed as the distance between the two stacks of D-branes and anti-D-branes.
The scalar field sector describing an  attraction between the pair of D3-brane and anti-D3-brane at long distance%
~\cite{HenryTye:2006uv,Dvali:1998pa,Dvali:2001fw,Burgess:2001fx}
has an effective coupling with the tachyon,
\begin{equation}
\mathcal{L}_{int}\, =\, -\lambda\, \phi^2\, T^2,
\end{equation}
where $\lambda$ is the coupling constant. 

\paragraph{The single tachyon cosmological model} 
The tachyon cosmology model was first proposed by~\cite{Sen:2002in, Sen:2002nu} and, independently, by Gibbons~\cite{Gibbons:2002md}. It depicts the picture that a pair of static D3-anti-D3 branes lay over each other
%%%%%
\footnote{ In the single tachyon field case, the dynamics of tachyon only describes the annihilation process of D-anti-D-brane pair, but does not include the issue that how these branes move to collide. To see how these branes move to collide with a weak attractive force between them, we will turn our attention to a coupled scalar-tachyon field model soon.}
%%%%%
and annihilate into closed string vacuum~\cite{Sen:2004nf}.  
In an effective field theory language,  the potential of the tachyon field has a maximum at $T=\dot{T}=0$.  During the annihilation process of the brane--anti-brane pair the tachyon field rolls down the potential hill and condenses. 
Right after the  tachyon condensation starts $\dot{T}\rightarrow1$ and $T\rightarrow\infty$ and the tachyon field behaves like cold matter
\begin{equation}
\rho_{T}=\frac{V(T)}{\sqrt{1-\dot{T}^2}}\propto a^{-3}~,\quad w_T=-\left(1-\dot{T}^2\right)=0~.
\end{equation}
This single tachyon field cosmological model has various applications, for example see~\cite{Gibbons:2002md, Fairbairn:2002yp, Feinstein:2002aj, Mazumdar:2001mm, Bagla:2002yn, Shiu:2002qe}. However, for the purpose of constructing bounce universe, such a lone tachyon does not suffice, since the tachyon's vacuum expectation value increases monotonically after condensation. 
In particular the universe driven by a single tachyon will contract to a cosmic singularity in a closed FLRW background~\cite{Sen:2003mv}.

%%%%%%%%%%%%%%%
\paragraph{The coupled scalar and tachyon  bounce (CSTB) model} 

In the presence of a canonical scalar  and  its coupling with the tachyon we take $\phi=\phi_0$ and $\dot{\phi}=T=\dot{T}=0$ as the initial conditions for the system. 

The picture of the CSTB model is that, at the  beginning,  
a stack of D3-branes and another stack of anti-D3-branes 
are separated by a long distance, $\phi=\phi_0$.  
The coupling term, $~\lambda T^2\phi_0^2$, stabilizes the system at $T=\dot{T}=0$. 
Due to a weak attractive force between D-brane and anti-D-brane ~\cite{HenryTye:2006uv,Dvali:1998pa,Dvali:2001fw,Burgess:2001fx}, 
modeled by the term, $-m^2_\phi\phi$, in the Lagrangian,
 the two stacks  will eventually encounter each other, $\phi\rightarrow 0$, and (some of the D-anti-D-brane pairs) annihilate into the closed string vacuum at the end of the tachyon condensation, $T\rightarrow \infty$. 
  
Furthermore, the CSTB model suggests a novel property: the vacuum expectation value acquired by the tachyon is finite 
but it never reaches infinite in our construct  (as shown in \refig{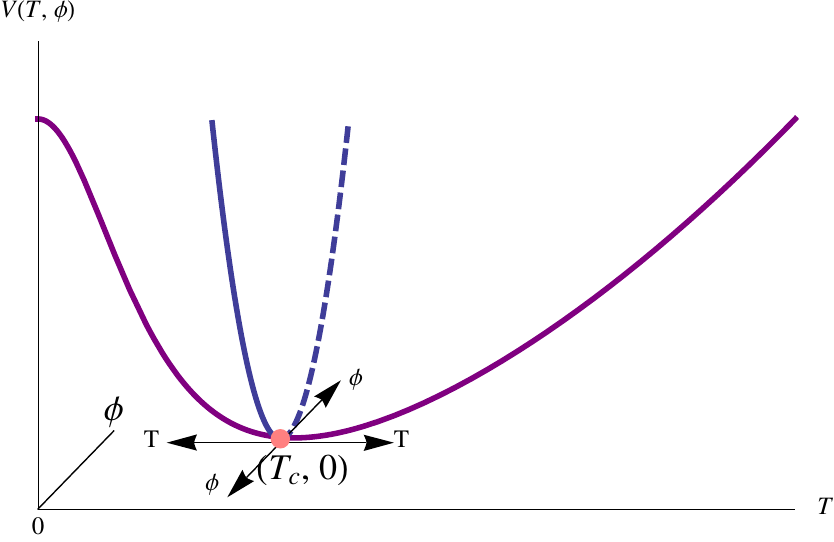})~\cite{Li:2011nj}.
 The tachyon  always gets pulled back and up the potential hill--due to the coupling with the scalar--before its condensation is completed. 
This property is key to the existence of the contraction--bounce--expansion cycles in the CSTB cosmos.  

Dynamically, the tachyon field and scalar field oscillate swiftly around $(T_c, 0)$ along the $T$-direction and the $\phi$-direction in $(T,\phi)$ field space, respectively,  with the commencement of the  tachyon condensation. 
During these oscillations, the average EoS of tachyon is
\begin{equation} \label{eq-awt}
\left\langle w_T \right\rangle=-\left(1-\langle\dot{T}^2\rangle\right)\simeq 0~,
\end{equation}
{\it{i.e.}} the tachyon field acts like a form of  cold matter once condensed, where $\langle * \rangle$ denotes the averaged value of the field  over a few  oscillations.

%%%%%%%%%%%%
%% figure
\fig{vacuum.pdf}{A sketch of  the effective potential of
the  scalar  and tachyon fields, $V(T, \phi)$, 
in the $(T,\phi)$ field space. 
The effective vacuum of CSTB cosmos is located at $(T,\phi)=(T_c,0)$. During the tachyon matter dominated phases of the CSTB cosmos, 
the tachyon  and the scalar  swiftly oscillate around 
$(T,\phi)=(T_c,0)$ along the $T$-direction and the $\phi$-direction, respectively.}{0.8}{h!t}
%%%%%%%%%%%%%

With~\refeq{totl}, one can study the cosmology of a universe governed by the coupled scalar-tachyon fields in the closed FLRW background. 
A cosmological solution with bounce/cyclic behaviour was obtained in~\cite{Li:2011nj}. 
One typical cycle  of the cosmological evolution comprises  the following three phases,
%%%%%%%%
\begin{enumerate}
\item {\bf Tachyon-matter-dominated contraction phase}%
%%%%%%
\footnote{~In the CSTB cosmos, as an auxiliary field, $\phi$ sector is always sub-dominated, and during this contraction, the averaged EoS of $\phi$ is also equal to zero, $w_\phi=0$. For simplicity, we call this phase as ``tachyon matter dominated contraction phase''.}% 
%%%%%%%
{\bf:} After the D3--anti-D3-brane  pair annihilate, the tachyon field condenses to ``tachyon matter,''  the  Equation of State(EoS) of which is equal to zero, $\langle w\rangle_T=0$. 
In a closed FLRW background the universe undergoes a 
matter-dominated contraction~\cite{Sen:2003mv, Li:2011nj}.  

\item {\bf The bounce phase:}  
A pair of D3--anti-D3-branes can be
pair-created, again, from the open string vacuum by vacuum fluctuations. 
The tension of these two branes behaves like a cosmological constant, $w_{branes}=-1$, 
the universe bounces smoothly from the pre-bounce contraction to a post-bounce expansion in the closed FLRW background%
%%%%%%%%%%%%%%%%%
\footnote{~One, if preferred, can picture a stack of D-branes and anti-D-branes, some of them undergo pair annihilation while some remain intact in each collision.}.
%%%%%%%%%%%%%%%%%
\item {\bf Post-bounce Expansion Phases:} After the bounce  the universe experiences an expansion driven by the tension of the branes preceding another expansion phase driven by the tachyon matter. 
One of these cycles can possibly  evolute into today's universe. 
\end{enumerate}
%%%%%%%%%%%%%%%%%%%%%%%%%

In  the bounce/cyclic universe scenario the primordial perturbations are generated and their subsequent exit of  the effective horizon all during the  pre-bounce contraction. 
To study the power spectrum of the primordial perturbations in CSTB cosmos  we, therefore, focus on the physics of tachyon-matter-dominated contraction. 

During the tachyon matter domination, according to the analytical results  and numerical simulations 
presented in~\cite{Li:2011nj},  the vacuum expectation value of tachyon field, $T_c$~, is proportional to the number of e-foldings 
of the cosmological background during contraction, $N_p$,
\begin{equation}
T_c\equiv\langle T \rangle\propto N_p~,\quad N_p\equiv\int H dt~.
\end{equation}
This is shown  in \refig{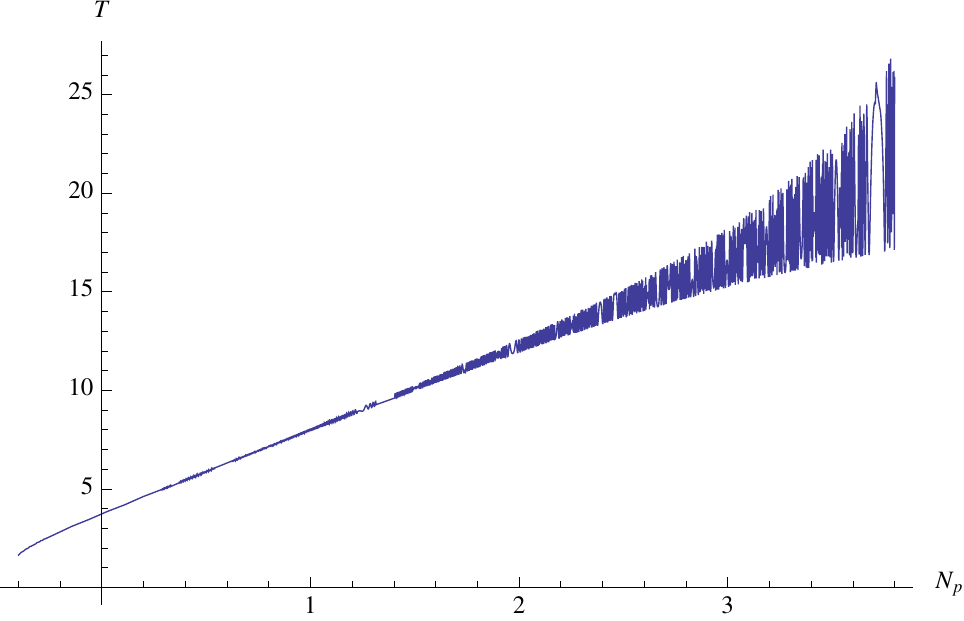} below.

%%%%%%%%%%%
\fig{Tcn.pdf}{A schematic plot of the evolution of the tachyon field  against the number of e-foldings, $N_p$, in the contraction phase ({\bf Right $\rightarrow$ Left}) in which the expectation value of tachyon field, $T_c$, evolves toward $0$. 
Notable is the linear dependence of 
$ \langle T_c \rangle$ on $N_p$.}{0.70}{h!t}
%%%%%%%%%%%%

Therefore $\dot{T}_c$ is linear in the Hubble parameter $H$,
\begin{equation}
\dot{T}_c=\kappa H~, \label{eq-tch}
\end{equation}
where we have decomposed $T_c$ as $T_c= \kappa N+\theta$, and both of $\kappa$ and $\theta$ are nearly constant. 
This property is, in turn, crucial to the successful generation of  scale invariant power spectrum of curvature perturbations in CSTB cosmos, to which  we will now turn our attention.

%%%%%%%%%%%%%%%
%%   Scale invariant Power Spectrum in CSTB Cosmos
%%%%%%%%%%%%%%% 
\section{Scale invariant Power Spectrum in CSTB Cosmos} 

To pave the road for the study of power spectrum of the primordial curvature perturbations generated by the tachyon field perturbations during the tachyon-matter-dominated contraction we derive the equations of motion for  $\delta\phi$ and  $\delta T$.
In Newtonian gauge~\cite{Mukhanov:1990me}, with $g_{\mu\nu}=diag \{-1-2\psi, a^2(1+2\psi))\delta_{ij}\}$
%%%%
\footnote{During the tachyon-matter-dominated contraction  in CSTB cosmos the  curvature term of FLRW metric, $K a^{-2}, K=1$, is  well sub-dominated. For simplicity we  use the flat FLRW metric and ignore tensor modes in the perturbation study.},
%%%%
we obtain
\begin{equation} \label{eq-dpe}
-\ddot{\delta \phi_k}+2\psi\ddot{\phi}-k^{2}a^{-2}\delta \phi_k+\left(-3H\dot{\delta \phi_k}-4\dot{\psi}\dot{\phi}+6H\psi\dot{\phi}\right)-\left(m^2_\phi+2\lambda T^2\right)\delta \phi_k=0~;
\end{equation}
\begin{eqnarray} \label{eq-dte}
&\ddot{\delta T_k}& -\, 2\, \psi\, \ddot{T}+k^2\, \delta T_k\, a^{-2}\, +\, (2\, \psi\, \dot{T}^2-2\dot{\delta T_k}\, \dot{T})\, \left(-\frac{1}{\sqrt{2}}+3H\dot{T}+2\lambda\phi^2T \frac{\sqrt{1-\dot{T}^2}}{V(T)}\right)\\ \nonumber
&+& (1-\dot{T}^2) \left[4\, \dot\psi\,\dot{T}\, +3H\, \delta\dot{T}\, -6\, H\, \psi\,\dot{T}\right]\\ \nonumber
&+& (1-\dot{T}^2)
\left[2\lambda\, \sqrt{1-\dot{T}^2}\, e^{T/\sqrt{2}}\, \phi\, \left(2\, T\, \delta \phi_k\, +\, \frac{\sqrt{2}+1}{\sqrt{2}}\phi\, \delta T_k + \psi\, \dot{T}^2 \, \phi -\dot{T}\, \phi\, \dot{\delta T_k}\right)\right]=0~,
\end{eqnarray}
%%%%%%%%%
where $\delta \phi_k$ and $\delta T_k$ are the Fourier modes of scalar and tachyon perturbations respectively.
%%%%%%%%%

According to~\refeq{dpe} the effective mass of $\delta\phi_k$, 
$M_{eff}=(m_\phi^2+2\lambda T^2)^{\frac{1}{2}}$, 
is very large during contraction and thus $\delta \phi$ is highly suppressed and can be safely neglected. Let us now turn our attention to the perturbations of the tachyon field. In general the background fields and its derivatives, 
which appear in \refeq{dte}, can be viewed as the external currents for the equation of motion for  tachyon's  perturbations. 
Taking the time-average of these fast-varying external currents \refeq{dte} can be simplified.  
During the contraction phase the background fields $\phi$ and $T$ oscillate swiftly around the effective vacuum of the system~\footnote{An analytic study of the dynamics of two coupled scalar fields can be found in~\cite{Wang:2011ed}.}, 
$(T,\phi)=(T_c,0)$ (as shown in \refig{vacuum.pdf}.
See~\cite{Li:2011nj} for  a detailed analysis.). 
Over a few complete  oscillations simplification is achieved because
\begin{equation}  \label{eq-ap}
\langle \phi\rangle=\langle \dot{\phi}\rangle=\langle \ddot{\phi}\rangle=0 
\end{equation}
and 
\begin{equation} \label{eq-at}
T_c\equiv\langle T\rangle~,\quad r_1\equiv \langle1-\dot{T}^2\rangle\simeq 0,\quad \langle \ddot{T}\rangle=0~.
\end{equation}
Furthermore  the ``effective driving force'' for the perturbations
\begin{equation}
r_2\equiv\left\langle-\frac{1}{\sqrt{2}}+3H\dot{T}+2\lambda\phi^2T \frac{\sqrt{1-\dot{T}^2}}{V(T)}\right\rangle\simeq0~, \label{eq-atf}
\end{equation}
in $T$-direction also vanishes~\cite{Li:2011nj}. 
Substituting \refeq{ap}, \refeq{at} and\refeq{atf} into \refeq{dte} we obtain a simplified equation of motion for the tachyon perturbations,
\begin{equation} \label{eq-pts}
\ddot{\delta T_k}+\frac{k^2}{a^2}\delta T_k=0~. 
\end{equation}
We have  taken $r_1 \sim r_2 \sim 0$ to the lowest order approximation;  but  we will put them back when we calculate the primordial spectral index.

Well before effective horizon exit at $|aH|\sim k$, each perturbation mode, $\delta T_k$, with wavenumber $k/a$ is evolving independently, and it is negligible at the classical level as the ``vacuum fluctuations''. 
However, after the effective horizon exit, $k\eta\rightarrow 0$, 
it grows to be a classical perturbation, which, in turn, determines the curvature perturbations evaluated on spatially flat slices. 
In the tachyon-matter-dominated contraction phase of CSTB cosmos, 
the cosmological background is evolving by a power-law,
\begin{equation}
a\propto \eta^2~,\quad \eta\rightarrow 0\label{eq-aeta}
\end{equation}
with $\eta$ being the  conformal time, $d\eta=a^{-1}dt$. 
Solving \refeq{pts} with \refeq{aeta} in the limit 
$k\eta\rightarrow 0$, we obtain the solution of each tachyon  perturbation mode after horizon exit
\begin{equation}
\delta T_k\propto k^{-\frac{3}{2}} \eta^0
\end{equation}
at leading order. 
And the power spectrum of tachyon perturbations becomes
\begin{equation} \label{eq-spt}
\mathcal{P}_{\delta T}\equiv\frac{k^3|\delta T_k|^2}{2\pi}\propto k^0\eta^0
\end{equation}
which is time-independent as well as  scale-invariant.

Turning  to Wands's model of matter dominated contraction,   the power spectrum of  primordial perturbation  can be re-casted into a simple form~\cite{Wands:1998yp,Li:2012vi}~,
$$
\mathcal{P}_{\delta \Phi}\propto k^0\eta^{-6}.
$$ 
During a perfectly  matter-dominated contraction as proposed by Wands, 
$a\propto \eta^2$ and $\eta\rightarrow 0$, 
the total energy density of the  cosmological background evolves as 
$\rho_b\propto \eta^{-6}$. 
The energy density  of the perturbations and  that of the background field grow with  the same rate, $\eta^{-6}$. 
However, in a realistic case that the cosmological background has a small departure from the perfectly matter-dominated contraction, $a\propto\eta^{2+\delta\nu}$ and $\delta \nu>0$, the energy density of field perturbations in a model like Wands's, 
$$
\rho_{\delta \Phi}\propto \mathcal{P}_{\delta \Phi}\propto\eta^{-6-4\delta \nu},
$$
grows faster than the energy density of the background field, 
$\rho_{\Phi}\propto \eta^{-6-2\delta \nu}$. 
It implies that the energy density of perturbations would become dominated during a contraction $(\eta\rightarrow 0)$ and henceforth rendering the cosmological evolution unstable
%%%
\footnote{A similar analysis and conclusion have been made in~\cite{Gratton:2003pe}. }.
%%% 
Moreover the time-dependence in the power spectrum of density perturbations derived from Wands's model also implies an implicit 
$k$-dependence when the perturbation modes exit the horizon at different moments, so that the power spectrum  is not truly scale-invariant.
 
To the contrary, the CSTB cosmos does not suffer these two problems. 
The time independence of power spectrum of CSTB cosmos 
\eqref{eq-spt} guarantee that the perturbations are always 
sub-dominated during a contraction phase and would not destabilize the background. 
Furthermore, perturbing the matter-dominated background, $a\propto\eta^{2+\delta\nu}$, the power spectrum of the tachyon field is still scale invariant and time independent, 
$\mathcal{P}_{\delta T}\propto k^0\eta^0$~.
Once again  the time independence ensures the power spectrum of tachyon field perturbations in the CSTB cosmos is explicitly scale invariant even through each perturbation mode exits the horizon at a different moment.
Therefore we conclude that the power spectrum of tachyon field perturbation in CSTB cosmos 
\eqref{eq-spt} is truly scale-invariant and stable under time evolution. 

\paragraph{Curvature perturbation of CSTB cosmos:}
To make  contact  with  observations we compute the power spectrum 
of  curvature perturbations evaluated on spatially flat slices in the  long wavelength limit~\cite{Senatore:2012tasi,Maldacena:2002vr},
\begin{equation}
\zeta=\frac{\delta a}{a}=H\delta t=\frac{H}{\dot{T}_c}\, \delta T~,
\end{equation}
where we have used the relation, 
$\delta T=\frac{d\langle T\rangle}{dt} \delta t=\dot{T}_c\, \delta t$. 
The power spectrum of curvature perturbations generated during tachyon-matter-dominated phase in CSTB cosmos then follows 
\begin{equation}  \label{eq-spct}
\mathcal{P}_\zeta=\left(\frac{H}{\dot{T_c}}\right)^2\mathcal{P}_{\delta T}~.
\end{equation}
With \refeq{spt} at hand we need to compute the $k$-dependence and time-dependence of the factor, $(H/\dot{T_c})^2$, in \refeq{spct} to determine whether or not the power spectrum of curvature perturbations  is stable and truly scale invariant in the cosmological sense. 
Using~\refeq{tch} we obtain the power spectrum of curvature perturbations
\begin{equation}
\mathcal{P}_\zeta=\kappa^{-2}\mathcal{P}_{\delta T}\propto k^0\eta^0.
\end{equation}
All in all we conclude that  the power spectrum of curvature perturbations  is stable and is cosmologically  scale-invariant.

\paragraph{Primordial spectral index:}
We will now present the computation of  the primordial spectral index of the curvature perturbations in the CSTB cosmos. 
In the last section  we take $r_1\,=\,r_2\,=\,0$ in the discussion of  scale-invariance of power spectrum at the lowest order. 
We shall hereby take their small values into account. 
By \refeq{dte} we obtain the equation of motion for the perturbations  of the tachyon,
\begin{equation}  \label{eq-dtns}
\ddot{\delta T_k}+\delta mH\dot{\delta T_k}+\frac{k^2_{e}}{a^2}\delta T_k=0~,
\end{equation}
where 
$\delta m=-r_12\dot{T}+r_2^2\left[2\lambda\, \rho_T \, \phi\, \left(  3H-\dot{T}\, \phi \right)\right] $, 
and 
$k_e^2\equiv k^2+a^2 m_e^2$ with $m_e^2$ 
being the effective mass of tachyon perturbations given by  $m_e^2=r_2^2 (2+\sqrt{2})\lambda \rho_T\, \phi^2$. 
Solving \refeq{dtns} in the cosmological background,
\begin{equation}
a\propto \eta^{2+\delta \nu}~,
\end{equation}
with $\delta \nu$ denoting the small deviation of CSTB cosmos background from a perfectly matter-dominated background, 
we obtain the power spectrum of tachyon field perturbation as
\begin{equation}
\mathcal{P}_{\delta T_k}=k^3k_e^{-3+2\delta m-\delta \nu}\eta^0~.
\end{equation}
Therefore the spectral index of curvature perturbation is 
\begin{eqnarray}
\nonumber n_s-1\equiv \frac{d \ln \mathcal{P}_\zeta}{d \ln k}=-2\frac{d\ln \kappa}{d\ln k}+\frac{d \ln \mathcal{P}_{\delta T_k}}{d \ln k}\qquad\qquad\\ 
=-2\frac{d\ln \kappa}{d\ln k}+2\delta m-\delta\nu+3\left(\frac{a^2m_e^2}{k^2}-\frac{d(a^2m_e^2)}{2k d k}\right)~. \label{eq-sindex}
\end{eqnarray}
The first term in the last line is from the factor relating curvature perturbations to  field perturbations. 
The value of the quantity, $\kappa$, is determined by the dynamics of background fields and principally independent of $k$. 
The second term relates the time-averaged quantities which nearly canceled during each oscillation cycle. 
The third term is derived from the small deviation of CSTB cosmos background from a perfectly matter-dominated background. 
And the last term indicates the influence  of the  effective mass on the tachyon field perturbations, which is negligible for the range of $k$ that we are interested in. 
All of these terms are not sensitive to the choices of initial conditions and the small changes in the shape of potential in CSTB cosmos. 
In other words, in contrast to the slow-roll inflation model's need of fine-tuning  the initial conditions and extreme flatness of potential to arrive at a small spectral index, the CSTB cosmos is much more stable as well as  natural in  having the value of $n_s-1$  around a few percents.

%%%%%%%%%%%%%%
%%%
%%%%%%%%%%%%%%
\section{CSTB cosmos {\it versus} Slow-roll Inflation} 

In this section we show the underlying ``dualities''  relating the slow-roll inflation~\cite{Liddle:2000cg,Dodelson:2003ft}, Wands's model~\cite{Wands:1998yp,Finelli:2001sr, Durrer:1995mz} and CSTB cosmos~\cite{Li:2011nj} in the 
$(\nu, m)$ parameter space. 
And we analyze the stabilities of the power spectra for these three models.

The equation of field perturbations of slow-roll inflation, Wands's model and CSTB cosmos--generally speaking--can be written as 
\begin{equation}
\ddot{\chi}_k+mH\dot{\chi}_k+\frac{k^2}{a^2}\chi_k=0~,\quad a\propto \eta^\nu~, \label{eq-gdet}
\end{equation} 
where $\chi_k$ denotes the Fourier mode of each field perturbation. For each model $(\nu, m)$ takes constant value in the parameter space:  
$(\nu, m)=(-1,3)$ for the slow-roll inflation, 
$(\nu, m)=(2,3)$ for Wands's model, and 
$(\nu, m)=(2,0)$ for the CSTB cosmos.
 
\paragraph{Parameter space of different cosmoses:}
In a model independent approach, solving \refeq{gdet}, one can obtain the power spectrum of $\chi_k$~,
 \begin{equation} \label{eq-gsp}
\mathcal{P}_\chi\equiv\frac{k^3|\chi_k|^2}{2\pi}\sim k^{2L(\nu,m)+3}\eta^{2W(\nu,m)}~,
\end{equation}
out of the effective horizon, $k\eta\rightarrow 0$~. 
The indices,  $L$ and $W$, are functions of $\nu$ and $m$:
\begin{equation}
L(\nu,m)=-\frac{1}{2}|(m-1)\nu-1|,~~~
W(\nu,m)=-\frac{1}{2}\{ [(m-1)\nu-1]+|(m-1)\nu-1| \}~.
\end{equation}
With \refeq{gsp} at hand we can obtain all scale-invariant solutions and time-independent solutions in the $(\nu, m)$ parameter space by solving the $k$-independence condition, $2L(\nu, m)+3=0$, and the time-independence condition, $W(\nu, m)=0$~. We  plot the solutions in the $(\nu, m)$ parameter space in~\refig{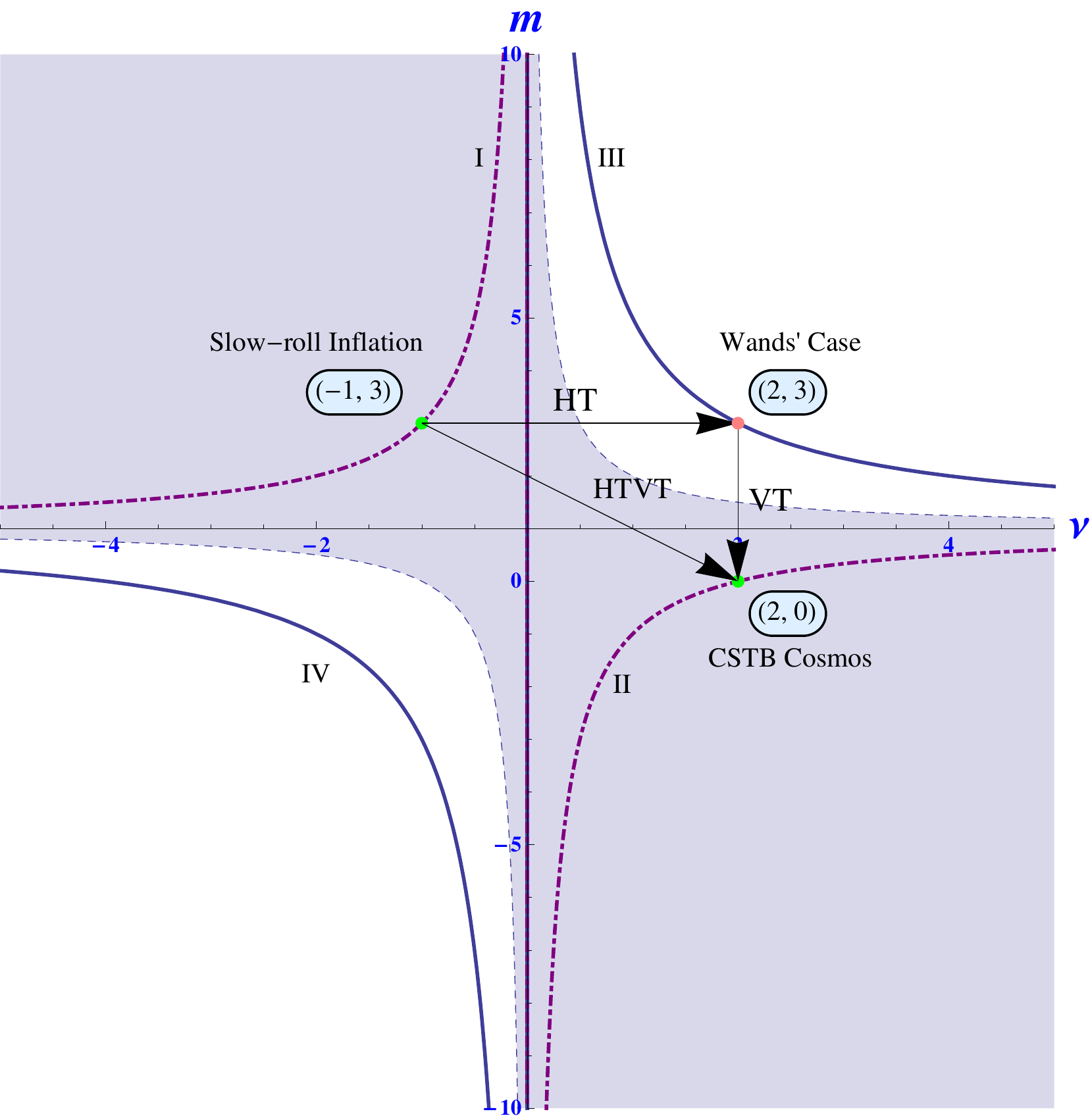}. 

\paragraph{Time-independent solutions:} To avoid the severe problem that  increasingly growing perturbation  modes may destabilize the cosmological background,  each  stable power spectrum of density perturbations should be time-independent, $W(\nu, m)=0$. 
In the $(\nu, m)$ parameter space, we find that all solutions satisfying
\begin{equation}
(m-1)\,\nu-1 < 0
\end{equation}
are time-independent, {\it i.e.} stable. 
In \refig{htvt.pdf}, the shaded region includes all time-independent solutions satisfying $(m-1)\,\nu-1 < 0$  whose boundaries are defined by $(m-1)\,\nu-1= 0$ and are drawn with thin dash lines. 
%%%%%%%%%%%%%%%%%%%%%%
\fig{htvt.pdf}{%
A  parameter space $(\nu,m)$ to classify the power spectra of density perturbations. 
$\nu$ is the power law index (horizontal axis) and 
$m$ is  the red/blue-shift index (vertical axis).
The shaded region includes all time-independent solutions satisfying $(m-1)\,\nu-1 < 0$  whose boundaries are defined by $(m-1)\,\nu-1= 0$ and are drawn with thin dash lines. 
The purple dot-dash lines obeying $(m-1)\,\nu =-2$ represent scale-invariant as well as time-independent solutions. 
Another set of scale invariant solutions given by  
$(m-1)\,\nu =4 $ (violet solid lines) have Fourier modes varying with time and therefore are not truly scale-invariant in a physical sense.%
}{0.8}{h!t}
%%%%%%%%%%%%%%%%%%%%%%%

\paragraph{Scale-invariant solutions:}
The scale-invariance condition,  $2L(\nu, m)+3=0$, yields four groups of scale-invariant solutions.  
Two of them, I and II, generated in expansion and contraction  respectively, are stable (time-independent),
\begin{equation}
\label{eq-ssis} 
\left\{  
  \begin{array} {l}
 {\displaystyle I: m = -\frac{2}{\nu}+1~, \quad \nu<0}   \\ 
 \\ 
  {\displaystyle II: m = -\frac{2}{\nu}+1~, \quad \nu>0}    \\
\end{array}     
\right. ~.
\end{equation}
In \refig{htvt.pdf} they are drawn with dot-dash lines. 
The other two groups, III and IV, also generated in expansion and contraction phase respectively, are unstable (time-dependent),
\begin{equation}
\label{eq-isis} 
\left\{  
  \begin{array} {l}
 {\displaystyle III: m = \frac{4}{\nu}+1~, \quad \nu>0}   \\ 
 \\ 
  {\displaystyle IV: m = \frac{4}{\nu}+1~, \quad \nu<0}    \\
\end{array}     
\right. ~.
\end{equation}
In \refig{htvt.pdf} they are drawn with solid lines.

In particular the slow-roll inflation, $(\nu, m)=(-1, 3)$~, and CSTB cosmos,  $(\nu, m)=(2, 0)$~, 
\begin{equation}
\mathcal{P}_{slow-roll}\propto k^0\eta^0,~~~\mathcal{P}_{CSTB}\propto k^0\eta^0~. \label{eq-slcs}
\end{equation}
belong to the class of  stable scale-invariant solutions, I and II, respectively. 
And the Wands's  model, $(m,\nu)=(3, 2)$~,
\begin{equation}
 \mathcal{P}_{Wands}\propto k^0\eta^{-6}~.
\end{equation}
belongs in the group of the unstable scale-invariant solutions, III. Now we turn our attention to the ``duality'' transformations which would connect these three models.

\paragraph{Duality transformations:}
In the $(\nu, m)$ parameter space shown in~\refig{htvt.pdf} 
there are two kinds of transformations connecting the stable scale-invariant solutions and unstable scale-invariant solutions, Horizontal Transformation (HT, relating cosmos of which perturbations with the same Hubble friction term but different background time evolution, {\it i.e.} ``iso-damping transformation" ), 
\begin{equation}
(m,\nu)\rightarrow (m, -\nu+\frac{2}{m-1} )~,
\end{equation}
and Vertical Transformation (VT, relating cosmos of which perturbations with different Hubble terms but same background time evolution, {\it i.e.} ``iso-temporal transformation"),
\begin{equation}
(m,\nu)\rightarrow (2-m+\frac{2}{\nu}, \nu )~.
\end{equation}
Under a  HT,  the two group of stable scale-invariant solutions, I and II~, are respectively mapped to the two group of unstable scale-invariant solutions, III and IV~, in the horizontal direction. 
Under a  VT, I and II are mapped to IV and III respectively in vertical direction. 

The solution of the slow-roll inflation, $(\nu, m)=(-1, 3)$~,  is connected horizontally  to Wands's Case, $(\nu, m)=(2,3)$~. 
Clearly this duality, which connects  Wands's matter-dominated contraction and slow-roll inflation shown in ~\cite{Wands:1998yp}
%%%%%%
\footnote{This duality is also discussed in various gauge choice with taking account in subdominated modes of perturbations~\cite{Boyle:2004gv, Piao:2004uq}},
%%%%%%
is a special case of all possible HT's with $m=3$. 
On the other hand, under a VT, the Wands's Case, $(\nu, m)=(2,3)$~, is dual to CSBT cosmos, $(\nu, m)=(2,0)$~. 

However, neither a HT nor a  VT is a complete operation since each of them only maps a stable scale-invariant solution to an unstable scale-invariant solution and {\it vice verse}.
 We are looking for a complete duality transformation connecting a stable scale-invariant solution in an expansion phase to another stable and scale invariant solution in a contraction phase. 
The simplest way is to perform  HT and VT consecutively,
\begin{equation}
(m,\nu)\xrightarrow{HT} (m, -\nu+\frac{2}{m-1} )\xrightarrow{VT} (2-m+\frac{2(m-1)}{-(m-1)\nu+2}, -\nu+\frac{2}{m-1}  )~,
\end{equation}
as shown in \refig{htvt.pdf}. 
We call this transformation a {\it{complete}} duality transformation (HTVT in \refig{htvt.pdf}). 
It can connect all stable scale-invariant solutions in an expansion  lying on  Line I to  Line  II of all possible stable and scale invariant solutions in a contraction.  
Interestingly the slow-roll inflation, $(\nu, m)=(-1,3)$~,  is related to the CSTB cosmos, $(\nu, m)=(2, 0)$, in this way--both possess  stable and scale invariant spectra  with the former generated in an exponential expansion while the latter in a tachyon-matter-dominated contraction.  

\paragraph{Stability Analysis:} 
Noting that both slow-roll inflation and CSTB cosmos can produce 
 power spectra of density perturbations satisfying current cosmological constraints. And they are ``dual'' to each other in the $(\nu, m)$ parameter space. 
However hidden in  their curvature perturbation spectra there is a significant difference in  fine-tuning.  
The curvature perturbations of slow-roll inflation model are
\begin{equation}
\mathcal{P}_{s-\zeta}=\left(\frac{H}{\dot{\Phi}}\right)^2\mathcal{P}_{\delta \Phi}~, 
\label{eq-rcfs}
\end{equation}
and those of the  CSTB cosmos are
\begin{equation}
\mathcal{P}_{c-\zeta}=\left(\frac{H}{\dot{T}_c}\right)^2\mathcal{P}_{\delta T}~, 
\label{eq-rcft}
\end{equation}
where $\mathcal{P}_{s-\zeta}$~ and $\mathcal{P}_{\delta \Phi}$~
being the spectra of the curvature perturbations and 
field perturbations for slow-roll inflation while those for 
CSTB cosmos being
$\mathcal{P}_{c-\zeta}$~ and $\mathcal{P}_{\delta T}$.
$\Phi$ and $T$ are the scalar field and  the tachyon field driving slow-roll inflation model and CSTB cosmos respectively. 

Given that $\mathcal{P}_{\delta \Phi}$ and $\mathcal{P}_{\delta T}$ are scale-invariant and stable%
%%%%%%%%%%%
\footnote{They only include the small derivations in their cosmological backgrounds from a perfectly exponential expansion or a perfectly  matter-dominated contraction respectively,
 which is not related to the fine-tuning problem. 
Therefore, we take them as perfect scale-invariant here for simplicity.},
%%%%%%%%%%%
to ensure the spectra of their curvature perturbations to be also stable and scale-invariant, both $\left(H/\dot{\Phi}\right)^2$~and $\left(H/\dot{T}_c\right)^2$ are required to be nearly constant. 
In slow-roll inflation models, 
to make $\left(H/\dot{\Phi}\right)^2$ nearly constant one needs to fine-tune the extreme flatness of the scalar potential.
 However in the case of   CSTB cosmos, $\dot{T}_c\propto H$,  is a dynamical attractor solution of the background field~\cite{Li:2011nj}.  
 According to \refeq{tch} 
 $\left(H/\dot{T}_c\right)^2\sim \kappa^{-2}$ is automatically--and always will be--nearly constant. 
Therefore we can conclude that the scale-invariance of curvature perturbation spectrum  in  the CSTB cosmos is more stable and free of fine tuning problem in contrast to that in  the slow-roll inflation model.

\section{Summary and Prospects}

In this paper we present a string-inspired bounce universe 
model--utilizing  the coupling of a canonical scalar with the ubiquitous string tachyon field to realize the bounce (CSTB cosmos). We obtain a spectrum of curvature perturbations that is  a stable and nearly scale-invariant. 
The  big bang singularity problem is resolved in CSTB cosmos universe by the 5-D  completion of  the D3-anti-D3-brane annihilation and creation processes~\cite{Li:2011nj}, which is no longer singular event when viewed from one dimension higher.

The pre-bounce contraction phase of CSTB cosmos--during which the cosmological perturbations are generated--is dominated by the condensing tachyon field, {\it i.e.} tachyon matter. 
Because of the Hubble friction term of tachyon field perturbation vanishes after tachyon condensation, $mH=3H\sqrt{\langle1-\dot{T}^2\rangle}\rightarrow 0$~, a time-independent and scale-invariant power spectrum of tachyon field is generated in this contraction phase. 

Furthermore, the power spectrum of curvature perturbation is related to that of tachyon field by a factor, $\left(H/\dot{T}_c\right)^2$, in long wavelength limit, $k\eta\rightarrow 0$~. 
According to the background evolution of CSTB cosmos~\cite{Li:2011nj}~, this  factor is a constant in both time and scale--independent of $k$ and $\eta$.  Therefore, the power spectrum of curvature perturbations is also stable and is scale-invariant in the cosmological sense.

We present a detailed study of the  spectral index of primordial curvature perturbations. We find that each term of this spectral index is insensitive to choices of initial conditions and/or the slight changes of cosmological background in CSTB cosmos. 
It  indicates that  CSTB cosmos is stable as well as natural in having the value of $n_s-1$ around a few percents consistent with observations. This may  serve as an explicit model for constructing a bounce universe in which the scale invariance of the power spectrum is generated during the pre-bounce contraction; and it is then preserved by the smooth bounce process in the long wavelength limit and subsequently  becomes the density perturbations of the obervable universe.  
 
To gain a deeper understanding of how  stable scale-invariance emerges  in the contraction phase of CSTB cosmos, 
we studied all scale-invariant and/or time-independent solutions in the unified parameter space of inflationary-bouncing cosmologies, {\it i.e.} the $(\nu, m)$ parameter space in~\refig{htvt.pdf}. 
We find a {\it complete} duality transformation--iso-background and followed by an iso-temporal transformation--connecting all stable scale-invariant solutions in an expansion to all stable scale-invariant solutions in a contracting phase. 
Interestingly the CSTB cosmos, $(\nu, m)=(2, 0)$, 
is related to  the slow-roll inflation, $(\nu, m)=(-1,3)$~, in this way--both possess stable and scale invariant spectra with the former generated in a tachyon matter dominated contraction while the latter in a well-known exponential expansion. 

Summing up our discussion, we note several issues for further studies. In this paper our study focuses on the generation and the evolution of perturbations during the pre-bounce contraction phase 
in CSTB cosmos.  Though it can be expected that the long wavelength modes of perturbations at the  classical level would not be perturbed significantly through the smooth bounce%
%%%%%%%%
\footnote{Recently many great progresses have  been made on studying the evolution of perturbations going through a bounce, for example, see~\cite{Deruelle:1995kd,  Finelli:2001sr, Allen:2004vz, Creminelli:2007aq, Lin:2010pf,Xue:2011nw}.}, 
%%%%%%%%
, a full investigation, of how these perturbations going through the smooth bounce and re-entering horizon in post-bounce expansion in CSTB cosmos, is a worthwhile exercise. 

On other hand, the perturbation spectra of slow-roll inflation and CSTB cosmos are (almost) identical, at the leading order,  in~\refeq{slcs}. 
To distinguish CSTB cosmos from the famous slow-roll inflation, the bispectrum of CSTB cosmos should be computed  to extract a specific prediction of the  shape of bispectrum in the CSTB cosmos. 
To conclude, we remark  that the unified parameter we introduced in this paper is not only useful for proving the stability and scale invariance of the slow roll and CSTB models, it also 
enlighten the search for such spectra from other early universe models.

\section{Acknowledgments}

Useful discussions with Robert Brandenberger,  Yifu Cai, Jin U Kang, Konstantin Savvidy, Henry Tye and Lingfei Wang are gratefully acknowledged. 

This research project has been supported in parts by the 
Jiangsu Ministry of Science and Technology 
under contract~BK20131264 
and by the 
Swedish Research Links programme of the Swedish Research Council (Vetenskapsradets generella villkor) under contract~348-2008-6049.

We also acknowledge 
985 Grants from the Ministry of Education, and the 
Priority Academic Program Development for Jiangsu Higher Education Institutions (PAPD).

\clearpage
\addcontentsline{toc}{section}{References}

\bibliographystyle{JHEP}

\bibliography{dual-reference}

\end{document}